\documentclass[11pt,twocolumn]{MyTightStyle}

\usepackage{graphics}
\usepackage{url}

\begin{document}

\title{2 P2P or Not 2 P2P?}

\author{
Mema Roussopoulos\\
\small{Harvard University, Cambridge, {MA}}
\vspace{.05in}\\
TJ Giuli\\
\small{Stanford University, Stanford, {CA}}
\and
Mary Baker\\
\small{HP Labs, Palo Alto, {CA}}
\vspace{.05in}\\
Petros Maniatis\\
\small{Intel Research, Berkeley, {CA}}
\and
David S. H. Rosenthal\\
\small{Stanford University Libraries, {CA}}
\vspace{.05in}\\
Jeff Mogul\\
\small{HP Labs, Palo Alto, {CA}}
}

\maketitle

\begin{abstract}
In the hope of stimulating discussion, we present a heuristic decision
tree that designers can use to judge the likely suitability of a P2P
architecture for their applications. It is based on the
characteristics of a wide range of P2P systems from the literature,
both proposed and deployed.

\end{abstract}

\section{\label{sec:Intro}Introduction}

Academic research in peer-to-peer (P2P) systems has concentrated largely
on algorithms to improve the efficiency~\cite{Stoica2001short},
scalability~\cite{Malkhi2002short}, robustness~\cite{Gummadi2003b},
and security~\cite{Wallach2002short} of
query routing in P2P systems, services such as indexing and
search~\cite{li2003short}, and dissemination~\cite{Kostic2003short}
for applications running on top of these
systems, or even all of the above~\cite{Kubiatowicz2000short}.  While
these improvements may be essential to enhancing the performance of some
P2P applications, there has been little focus on what makes an
application ``P2P-worthy,'' or on which other, previously ignored
applications may benefit from a P2P solution.  What questions should an
application designer ask to judge  whether a P2P solution is
appropriate for his particular problem?

In this position paper, we hope to stimulate the discussion by
distilling the experience of a broad range of proposed and deployed
P2P systems into a methodology for judging how suitable a P2P
architecture might be for a particular problem.  In
Section~\ref{sec:characteristics}, we identify some salient
characteristic axes in typical distributed problems.  In
Section~\ref{sec:problems}, we describe a spectrum of specific
problems for which P2P solutions have been proposed.  In
Section~\ref{sec:decision}, we propose an arrangement of problem
characteristics into a heuristic decision tree.  We walk through the
tree explaining its choices and why we believe certain paths may lead
to successful P2P solutions to important problems, while other paths
may encounter difficulties.

\section{Characteristic Problem Axes}
\label{sec:characteristics}

In this section, we describe the characteristics of distributed
problems we believe are important in assessing their P2P-worthiness.
As seems to be the consensus in the research
community~\cite{Milojicic2002short}, we identify as peer-to-peer those
environments that satisfy the following three criteria:
\begin{itemize}
\item \emph{Self-organizing}: Nodes organize themselves into a network
through a process in which they discover other peers.
\item \emph{Symmetric communication}: Peers are considered equals; they
both request and offer services, rather than being confined to either
client or server roles.
\item \emph{Decentralized}: There is neither a global
directory nor a central controller dictating behavior to individual
nodes.  Instead, peers rely on communication with each other for
discovery of other peers, resources and services, and determine their
course of action autonomously.
\end{itemize}

Our axes are the problem's budget, the
relevance of resources to individual peers, the rate of
system change, the level of mutual trust, and the criticality of the
resources handled.  In more detail:

{\bf Budget}: If the budget for a solution is ample, a designer is
unlikely to consider worthwhile the inefficiencies, latencies and
testing problems of a P2P solution.  If the budget is limited, a key
motivator in the choice of P2P architectures is the lowest possible
cost of entry for individual peers, despite increased total system
cost.  Assembling a system from local, often surplus, components can
be justified as a small part of many budgets and may be the only
feasible approach.

{\bf Resource relevance to participants}: Relevance is the probability that
a peer is interested in data from other peers.  If it is high,
P2P cooperation evolves naturally.
If it is low,
artificial or extrinsic incentives may be needed.

{\bf Trust}: Mutual distrust between peers may be essential to the
problem or of little concern. However, the cost of mutual distrust in
P2P systems is high and needs to be justified by specific problem
requirements.

{\bf Rate of system change}: The participants, resources and
parameters of the system may be stable or rapidly changing.  Rapid
change in P2P systems can make it difficult to provide consistency
guarantees and defenses against flooding and other attacks.

{\bf Criticality}: If the problem being solved is critical to the users,
they may demand centralized control irrespective of technical criteria.
Even if P2P is not ruled out,
the need for expensive security or massive over-provisioning may make
it uneconomic.

We have excluded other characteristics which, while important,
did not affect the decision tree as far as we have elaborated it.
Among these are whether the resources are public or private, whether the
resources are naturally distributed, whether the time horizon of the
application is long or short, and whether participants are homogeneous
or heterogeneous.

\section{Candidate Problems}
\label{sec:problems}

Our candidate problems for a P2P architecture come from
routing, backup, monitoring, data sharing, data dissemination, and auditing.

\subsection{Routing Problems}
\label{sec:problems:routing}

All distributed systems need a routing layer to get messages to their
intended recipients.
Routing takes on P2P characteristics when the scale is large enough
(e.g., the Internet) or when centralization is ruled out 
(e.g., wireless ad hoc networks).

\subsubsection{Internet Routing}

Internet routers must communicate to cope with dynamically changing
network topology to determine how to route outbound packets to their
destination.  They are arranged into ``autonomous systems'' which
``peer'' with each other across organizational boundaries,
frequently between competitors.

Routing protocols have historically assumed that economic incentives
and legal contracts are sufficient to discourage misbehavior.
At the application layer (e.g., Resilient Overlay Networks
(RON)~\cite{Andersen2001short}) or at the network layer (e.g.,
BGP~\cite{BGP}), routers trust information from known peers.  They
cooperate because the information being exchanged is of interest to
all peers and important to their function.  This cooperation tends to
fail if error, misbehavior or usage patterns cause the data to change
too fast.  To scale to the size of the Internet, BGP tries to limit
the rate of change by aggregating routes instead of having ISPs
propagate internal routing updates.  Aggregation reduces the ability
to detect path outages quickly~\cite{Labovitz2000short}.  RON instead
gives up scaling to large numbers of nodes in favor of more
fine-grained route information exchanges.

\subsubsection{Ad hoc Routing in Disaster Recovery}

The ad hoc routing problem is to use transient resources, such as the
wireless communication devices of a disaster recovery crew, to deploy
temporary network infrastructure for a specific purpose.  Because each
individual node's wireless range does not reach all other nodes, peers
in the network forward packets on behalf of each other.  The costly
alternative is to provide more permanent infrastructure for all
possible eventualities in all possible locations.  The network is of
relevance and critical to all participants, and pre-configured
security can give a high level of mutual trust.  Once established, the
participants (humans in the crew) typically change and move slowly, and
do not exchange huge volumes of data.

\subsubsection{Metropolitan-area Cell Phone Forwarding}

Ad hoc routing has also been proposed in less critical settings, such
as that of public, ad hoc cellular telephony in dense metropolitan
areas.  The motivation is to reduce the costs of deploying enough base
stations and to avoid payment for air time where traffic does not pass
through base stations.  Unlike the disaster recovery problem, the
participants do not trust each other, they change and move rapidly,
and their local resources such as battery power are limited.  In its
current state, this problem suffers from the ``Tragedy of the
Commons"~\cite{Hardin1968}.  We doubt that a practical P2P solution to
this problem exists, unless either on-going
research~\cite{Bansal2003short,Buttyan2003short} devises strong,
``strategy-proof'' mechanisms to combat selfishness, or the scope of
the problem is limited to close-knit communities with ``built-in''
incentives for participation.

\subsection{Backup}

Backup, the process in which a user replicates his files in
different media at different locations to increase data survivability,
can benefit greatly from the pooling of otherwise underutilized
resources.
Unfortunately, the fact that each peer is interested only in its
own data opens the way to selfish peer behavior.

\subsubsection{Internet Backup}

The cost of backup could be reduced if Internet-wide
cooperation~\cite{Cox2003short,dabek01short} could be incentivized and
enforced.  For example in Samsara~\cite{Cox2003short} peers must hold
real or simulated data equivalent to the space other peers hold for
them.  But there is no guarantee an untrusted node will provide backup
data when requested, even if it has passed periodic checks to ensure
it still has those data.  Such a misbehaving or faulty node may in
turn have its backup data elsewhere dropped in retaliation.  If
misbehaving, it may already have anticipated this reaction and, if
faulty, it is exactly why it would participate in a backup scheme in
the first place.  We believe that data backup is poorly suited for a
P2P environment running across trust boundaries.

\subsubsection{Corporate Backup}  

In contrast, when participants enjoy high mutual trust, e.g., within the
confines of an enterprise intranet, P2P backup makes sense
(HiveCache~\cite{HiveCache} is one such commercial offering).  This is
because selfish behavior is unlikely when a sense of trusting community
or a top-down corporate mandate impose participation, obviating the need
for enforceable compliance incentives.

\subsection{Distributed Monitoring}

Monitoring is an important task in any large distributed system.  It
may have simple needs such as ``subscribing'' to first-order events
and expecting notification when those events are ``published'' (e.g.,
Scribe~\cite{Rowstron2001Scribeshort}); it may involve more
complicated, on-line manipulation, for instance via SQL queries, of
complex distributed data streams such as network packet traces, CPU
loads, virus signatures (as in the on-line network monitoring problem
motivating PIER~\cite{Huebsch2003short}); it may be the basis for an
off-line, post mortem longitudinal study of many, high-volume data
streams, such as the longitudinal network studies performed by Fomenkov
et al.~\cite{LongitudinalCaidashort}.

Although the abstract monitoring problem is characterized by natural
distribution of the data sources monitored, specific instances of the
problem vary vastly.  A longitudinal off-line network study, though
important, is not necessarily critical to its recipients, and has low
timeliness and rate-of-change constraints.  In contrast, an ISP may
consider the on-line, on-time monitoring of its resources and those of
its neighbors' extremely critical for its survival.  Similarly, the
mechanisms for complex network monitoring described by Huebsch et
al.~\cite{Huebsch2003short} may be appropriate for administratively
closed, high-trust environments such as PlanetLab, and they may be quite
inappropriate in environments lacking mutual trust and rife with fraud
or subversion; whereas an off-line long-term network study affords its
investigators more time for data validation against tampering.

\subsection{Data Sharing}

\subsubsection{File sharing}

In file sharing systems, participants offer their local files to other
peers and search collections to find interesting files.  The cost of
deployment is very low since most peers store only items that they are
interested in anyway.  Resource relevance is high; a great deal of
content appeals to a large population of peers.  In typical file sharing
networks, peer turnover and file addition is high, leading to a high
rate of change of the system.  Peers trust each other to deliver the
advertised content and most popular file sharing networks do not have
the capacity to resist malicious peers.  File sharing is mainly used to
trade media content, which is not a critical application.

\subsubsection{Censorship Resistance}

The goal of the FreeNet project~\cite{Clarke2000short} is to create an
anonymous, censorship-resistant data store.  Both publishing and
document requests are routed through a mix-net and all content is
encrypted by the content's creator.  These steps are necessary because
peers are mutually suspicious and some peers may be malicious.  Peers
share their bandwidth as well as disk space, which puts FreeNet on the
low end of the budget axis.  FreeNet is intended to provide a medium for
material that some group wishes to suppress, thus data are relevant to
many consumers as well as attackers.  It is critical that the content in
the system be protected from censorship.  Published material does not
need to be available immediately, so FreeNet can work with a low rate of
change.

Tangler~\cite{mazieres01short} has similar goals.  A peer stores his
document by encoding the document using erasure codes and distributing
the resulting fragments throughout the community.  To prevent an
adversary from biasing where those fragments are distributed, a peer
must combine his document with pseudo-randomness before erasure
coding; he uses other peers' documents as a source of
pseudo-randomness.  To retrieve his own document, a peer must store
locally the randomness used at storage time, i.e., other peers'
documents.  Although the problem lacks inherent incentives for
participation, this solution ingeniously supplies them.

\subsection{Data Dissemination}

Data dissemination is akin to data sharing, with the distinction that
the problem is not to \emph{store} data indefinitely but, instead, to
\emph{spread} the data for a relatively short amount of time.  Often
storing is combined with spreading.

\subsubsection{Usenet}

Usenet, perhaps the oldest and most successful P2P application, is a
massively distributed discussion system in which users post messages to
``newsgroups.''  These articles are then disseminated to other hosts
subscribing to the particular newsgroup, and made available to local
users.  Usenet has been a staple of the Internet for decades, arguably
because of the low cost barrier to peer entry and the high relevance of
the content to participating peers.  Unfortunately, although the system
flourished at a time when mutual trust was assumed, it remains
vulnerable to many forms of attack, perhaps jeopardizing its future in
less innocent times.

\subsubsection{Non-critical Content Distribution}

Dissemination of programs,
program updates, streaming media~\cite{Cohen2003short,Kostic2003short},
and even cooperative web caching~\cite{Wolman1999short} are all
non-critical applications of content distribution.  The
problem involves content with generally low rates of change,
although the participants may change wildly.

One successful application is BitTorrent~\cite{Cohen2003short}, which
mitigates the congestion at the download server when a popular new
program or update is posted.  Its tit-for-tat policy is effective
despite low peer trust, and the option of postponing download until
later reduces its criticality.

Collaborative web caching, although superficially attractive, has not
succeeded.  As the system scales up, relevance of the content
decreases, making it less useful.  When the scale is small enough to
make the content relevant, the system's complexity is
unjustified~\cite{Wolman1999short}.

\subsubsection{Critical Flash Crowds}

Other specific instances of dissemination have been proposed to
address flash crowds~\cite{Stading2002short,Stavrou2002short} which
could be used to distribute critical data, such as news updates during a
major disaster.

\subsection{Auditing}

\subsubsection{Digital Preservation}

The \emph{LOCKSS} system preserves
academic e-journals in a network of autonomous web caches.  Peers each
obtain their own complete replicas of the content by crawling the
publisher's web site.  If the content becomes unavailable from the
publisher, the local copy is supplied to local readers.  The replicas are
preserved using a P2P protocol~\cite{Maniatis2003lockssSOSPshort} that
provides mutual audit and repair, but this is not time critical; thus
the rate of change is low.  The content being preserved is highly
relevant to many peers.  The audit process uses ``opinion polls'' so
that peers trust the consensus of other peers but not any individual
peer.  Mutual distrust is essential to prevent cascade failures which
could destroy every copy of the preserved content.  The automatic audit
and repair process allows peers to be built from cheap, unreliable
hardware with very little need for administration, which is
important in the budget-constrained world of libraries.

\subsubsection{Distributed Time Stamping}

A secure time stamping service~\cite{Haber1991short} acts as the digital
equivalent of a notary public: it maintains a history of the creation
and contents of digital documents, allowing clients who trust the
service to determine which document was ``notarized'' first.
Correlating the histories of multiple, mutually distrustful secure time
stamping services~\cite{Maniatis2002bshort} is important, because not
everyone doing business in the world can be convinced to trust the same
centralized service; being able to map time stamps issued elsewhere to a
local trust domain is essential for critical documents (such as
contracts) from disparate jurisdictions.  Luckily, sensitive documents
such as contracts change little or not at all at very low rates, and
high latencies for obtaining or verifying secure time stamps are
acceptable, facilitating the development of an \emph{efficient-enough}
P2P solution to the problem.

\section{2 P2P or Not 2 P2P?}
\label{sec:decision}

\begin{figure*}
\centerline{\includegraphics{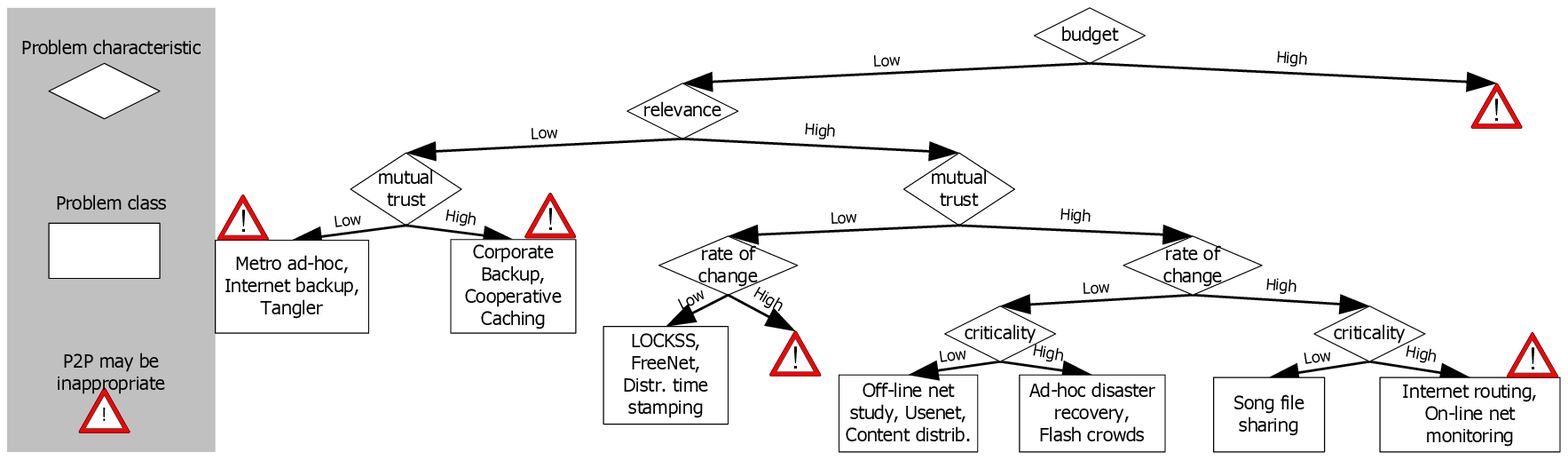}}
\caption{A decision tree for analyzing the suitability of a P2P solution.}
\label{fig:DecisionTree}
\end{figure*}

Figure~\ref{fig:DecisionTree} is a decision tree organizing
our characteristics to determine whether a particular application is
P2P-worthy.
We examine our example applications traversing this tree in a breadth-first
manner.

At the top of the tree we have the ``budget'' axis.  We believe that
limited budget is the most important motivator for a P2P solution.
With limited budget, the low cost for a peer to join a P2P solution is
very appealing.  Otherwise, a centralized or centrally controlled
distributed solution will provide lower complexity and higher
performance for the extra money.  Our tree thus continues only along
the ``limited'' budget end of the axis.

Our next most important characteristic is the ``relevance'' of the
resource in question.  The more relevant (important to others) the
resource, the more motivated peers in a P2P architecture are to
participate.  Applications of low relevance with good P2P solutions
exist, but only where other inherent reasons for a P2P solution are
strong, as we explain below.

The next axis in the tree is ``mutual trust.''  Successful P2P
solutions with trusting peers exist, as do those whose other
characteristics justify the performance and complexity cost of
measures to cope with mutual distrust.  Those applications with low
relevance and low trust have the burden of incentivizing cooperation.
While Tangler is a good example of this, we believe that metropolitan
ad hoc wireless networks and Internet backup have not yet succeeded.
The motivations for these applications seem inadequate to overcome the
low relevance of the resources and the overheads of protecting against
uncooperative or malicious peers.  Where peers are assumed to
cooperate, applications such as corporate backup may succeed, since
corporate mandate compensates for low relevance.  Unfortunately, no
such external compensation appears to exist for cooperative web
caching; its marginal performance benefits due to low relevance at large
scales renders it unnecessary~\cite{Wolman1999short}.

Where relevance is high, the level of trust between nodes still has an
impact on the suitability of a P2P solution for the application.
Creating artificial economies or ``trading'' schemes to provide
extrinsic incentives for cooperation (as in MojoNation) is generally
unsuccessful~\cite{McCoy2002short}.  The overhead in terms of
complexity and performance for managing mutually distrustful nodes
suggests that applications will be difficult to implement successfully
in a P2P system, unless other characteristics intercede to simplify
the problem.

Such a characteristic is the rate of change in the system.
Applications with a low rate of change, such as LOCKSS, FreeNet, and
distributed time stamping, may succeed despite mutually distrusting
peers.  For these applications, mutual distrust between peers is an
inherent part of the problem, and thus its cost must be born by any
proposed solutions.  The cost, however, is reduced by the low rate of
change, which makes it possible to detect problems in the system in
time to address them, and reduces the performance impact of the
measures to protect against malicious peers.  P2P applications with a
high rate of change in untrustworthy environments are unlikely to
succeed.

The rate of change in the system remains important even for
applications in which peers may trust each other to cooperate.  If the
rate of change is low, then both non-critical applications (such as
off-line network studies, Usenet, and content distribution) and
critical applications (such as ad hoc wireless network deployment for
disaster recovery and flash crowd mitigation) may succeed.  If the
system moves quickly, we believe that it is easier to deploy
non-critical applications such as song file sharing.  When the problem
involves critical information that also changes quickly (as in the
case of Internet routing and on-line network monitoring), the designer
should consider whether the application benefits sufficiently from
other features.  To the degree that Internet routing is successful, it
is because it is amenable to trading accuracy for scalability through
techniques such as aggregation of data.  If network monitoring
succeeds, it will be because the natural distribution and high volume
of the data allow few other appropriate solution
architectures beyond P2P.

\section{Conclusions}

To summarize, the characteristics that motivate a P2P solution are
limited budget, high relevance of the resource, high trust between
nodes, a low rate of change in the system, and a low criticality
of the solution.  We believe that the limited budget requirement is
the most important motivator.  Relevance is also very important but
can be compensated for by ``saving graces'' such as assumed
trust between nodes or strong imposed incentives.
Lacking these,
we believe that applications of low relevance are not
appropriate for P2P solutions.  Trust between nodes greatly eases P2P
deployment, however there are some applications, such as LOCKSS,
FreeNet and distributed time stamping, where deployment across trust domains is an
inherent requirement.  These applications must pay the overhead of
distrust between nodes, but are feasible in a P2P context because
a low rate of change makes these costs manageable.

While P2P solutions offer many advantages, they are inherently complex
to get right and should not be applied blindly to all problems.  In
providing a framework in which to analyze the characteristics of a
problem, we hope to offer designers with some guidance as to
whether their problem warrants a P2P solution.

\section{Acknowledgments} 

This work is supported by the NSF (Grant No.\ 9907296), by Sun
Microsystems Laboratories, by DARPA (contract No.\ N66001-00-C-8015),
and by MURI (award No.\ F49620-00-1-0330).


\end{document}